# Emergence of Dynamical Coherence in a Driven One-dimensional Interacting Rotor Model

**Sahithya S. Iyer**[§,1], **Sayantan Mondal**[†,1], *and* **Biman Bagchi**[†,*]

[§] Molecular Biophysics Unit, Indian Institute of Science, Bengaluru

[†] Solid State and Structural Chemistry Unit, Indian Institute of Science, Bengaluru

Karnataka 560012, India

*Corresponding author's email: bbagchi@iisc.ac.in

## Abstract

In order to understand the dynamics of active matter, we examine a minimalistic model where interacting spins on a one-dimensional lattice are driven by a self-propelled spin at the centre with a fixed rotational velocity ($\omega_0$). The other spins execute rotational Brownian motion by following the Shore-Zwanzig model of rotational dynamics. The simplicity of the model allows us to inquire about several relevant microscopic quantities. The continuous 'active' torque on the central spin is propagated through nearest neighbour interactions with a uniform coupling parameter, *J*. We have found a *bounded region* in the *J*-$\omega_0$ plane where the system exhibits 'active matter like behaviour'. Interestingly, in the limits of large *J* and $\omega_0$, we observe a 'slipping behaviour'. The site specific average rotational velocity of the spin, as one moves away from the central spin exhibits a nearly exponential decay with distance, allowing the definition of a correlation length ($\xi$) which grows rapidly with an increase of the coupling (*J*) between the spins. Site specific average velocity exhibits a change from a single exponential to biexponential decay pattern as the system enters the active region of the phase diagram, accompanied by a non-monotonic behavior of the correlation length. We conclude that a macroscopic coherent state can emerge in the presence of a small concentration of active molecules. We discuss experimental relevance of our results.

## 1. Introduction

Since the pioneering work by Vicsek *et al.*[1] a large number of studies have been devoted to understand the statistical mechanics of active matter.[2-12] This is naturally a subject of great relevance to biological systems where one often finds a coherent pattern of behaviour to emerge from seemingly uncorrelated motions.[13-15] Several pioneering theoretical studies have elucidated how standard hydrodynamic models can be generalized to include the motion in active matter, leading in turn to an elegant description of the emergence of a dynamically ordered state.[3, 16, 17]

---

[1] S.S.I. and S.M. contributed equally.

The Vicsek model is the most prominent example of a minimal model which is capable of capturing the physics of dynamical ordering transition exhibited by active particles.[1, 18, 19] The velocity-velocity interactions in this two-dimensional model results in a phase transition from a disordered state to a long-range correlation in the velocity of the driven particles as the noise (an equivalent of temperature) in the system decreases and system density increases. The emergence of order in the limiting case of low noise and high density has also been observed in simulation studies of one-dimensional Vicsek model.[20] The hallmark of Vicsek model is the emergence of a dynamically ordered state via a phase transition in two-dimensions. Ordered state is marked by the particles that move in the same direction and exhibits a 'flocking' motion.

We note in passing that in a strict sense an order-disorder phase transition cannot exist in an equilibrium two-dimensional systems as stated in the Mermin-Wagner theorem.[21] However, such restrictions do not exist in nonequilibrium systems. Activity induced phase transtition has also been reported in simple colloidal systems.[22] Earlier studies have also used the Péclet number (that controls the self-propulsion) to monitor the activity of articles in the system to induce activity controlled phase transition.[23] A recent study connects the relation between phase transition and the emergence of collective motion in a one-dimensional spin system.[24]

It is important to note the absence of explicit intermolecular interactions in the Vicsek model. This model achieves coherence state only by the application of a conditional 'velocity update rule' where particles move synchronously with their neighbours with deviations in a permissible noise range. More recently, there have been attempts to include intermolecular interactions among the participating particle, more in the spirit of the liquid state theory.[25, 26] The advantage of such a microscopic approach is that one can include molecular length scale processes such as collisions. Also, one can attempt to understand how microscopic time scale transcends into macroscopic time scales in a sense similar to the renormalization group theory approach to the critical phenomena. This is a formidable aim and the development can be regarded at an initial stage.

In several earlier papers, interaction between active particles has been included through a hydrodynamic advective term.[3, 5] We refer to the elegant work by Ramaswamy which showed how such an interaction can give rise to a motion in a given direction, thereby achieving a more physical realization of active matter than provided by the Vicsek model.

An important aspect of an active matter is the emergence of a flowing state which is a dynamically ordered state. This is a low entropy ordered state that bears certain similarity to Progogine's dissipative structure.[27] Of course Prigogine's dissipative structures do not have a flow built in. Nevertheless, the seminal aspect of the latter work is the appearance of a dynamically stable state characterized by minimum rate of entropy production. This provides a variational principle that allows selection of the preferred state.

In a recent paper, Vaikuntanathan and co-workers[28] studied the evolving structure, dynamics and fluctuations in a system of two-dimensional disks subject to three kinds of forces. First, each disk experiences a conservative force due to distance-dependent repulsive interaction forces among themselves, as in any standard liquid state theory. Second, the disks experience a stochastic random force that obeys Gaussian statistics and gives rise to dissipation. It is the third force, whose form is motivated by an experimental active colloid mixture study,[29] that simulates active matter. This is an external force acting on half of the particles, absent for the rest. This external force changes direction periodically. The external force is so chosen that in the zero-temperature limit, a single externally driven particle will trace a circle in the XY plane. The periodically changing external force induces a coexistence between the phases characterized by work done. Work done is defined differently as the energy dissipation.[30-32] Diffusion constant of the system, controlled by the amiplitude of non-conservative external forces, increases due to energy dissipation and determines properties of the non-equilibirum system. There are regions where energy dissipation is more and similar other regions where dissipation is less. These factors control the phase separation behaviour of the non-equilibrium system.

It is interesting to note that several years ago, several studies considered the microscopic structure of a liquid made of spherical molecules (like colloidal spheres) subjected to shear flow.[33-35] In particular, considerable attention was devoted to the melting of colloidal crystals under shear flow. It was observed that the structure of a liquid under shear flow develops anisotropy in the manner that the structure factor in the direction of flow becomes different from those in the perpendicular direction. This altered structure influences the freezing-melting behaviour of the system.[36] This bears resemblance to the present study of active matter under flow once a steady state is reached.

Our motivation is to study the emergence of a dynamically coherent steady state in a driven Hamiltonian system. In particular, we are interested in investigating the possible modification in the microscopic correlations in a nonequilibrium steady state in an interacting system. As the spins interact with each other, we have the scope of studying the emergence of a coherent steady state as a function of the interaction potential and the driving force. We have indeed observed the formation of such a steady state far from equilibrium.

Our study of active matter that is based on one-dimensional XY like spin system. In this paper, we explore how the controlled self-driven rotation of a spin at the centre of the one-dimensional chain affects the natural Brownian motional of all other spins. The spins are coupled by nearest neighbour interactions. This is perhaps the simplest conceivable model of interacting systems, much in the same spirit of the Glauber model that played an important role in modelling the dynamics of interacting systems.[37-40] However, we needed to employ a continuum version of the Glauber model, and the model of choice is the celebrated Shore-Zwanzig model which allows not only but also Brownian diffusion of the system of interacting spins.[41]

In the present study we attempt to answer the following questions: How does one describe the effects of one spin driving the rest of the system through nearest neighbour interactions? This is clearly the scenario when information and effects propagate through interactions. In the presence of a driving force correlations appear which under weak coupling nevertheless decays as the effects diminish with distance. However, the length scale of the decay of correlations remains an interesting point to explore. The second question is the possible emergence of a coherent state where all the spins rotate in unison with the central driving spin. This is thus a state of zero entropy, if we could define entropy in terms of the width of distribution of velocities.

2. **Description of the Model**

The model, as shown in **Figure 1**, consists of a one-dimensional lattice of spins where each spin is free to rotate about its axis. The central spin (#0) is driven externally and rotates with a constant velocity $\omega_0$. The spins interact with their nearest neighbours following the Shore-Zwanzig Hamiltonian and the entire system of spins undergoes continuous Brownian diffusion motion. Shore-Zwanzig Hamiltonian is similar to the classical XY model where the spins have continuous rotational degree of freedom in the XY plane.[41, 42] In the absence of a uniform external potential acting on the system, the interaction Hamiltonian is given by Eq.(1).

$$H = -J\sum_{\langle i,j \rangle} \cos(\theta_i - \theta_j) \qquad (1)$$

where the indices i and j are nearest neighbour sites and *J* is the coupling strength parameter. When there is no external driving force on the spins, at equilibrium and lower temperatures, the spins attain the same orientation. In the case of driven systems, when the central spin rotates at a constant velocity $\omega_0$, the interaction between spins, governed by the coupling constant in Shore-Zwanzig Hamiltonian, determines the spatial propagation of the velocity from the central spin. Above a given coupling strength, there emerges a coherent motion.

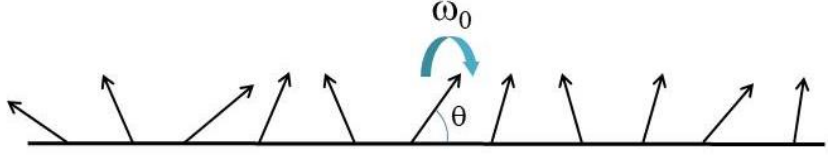

**Figure 1. A schematic illustration of the model: Spins arranged in a 1D lattice with the central spin rotating at a constant velocity of ω₀. The state of each spin is uniquely defined by an angle (θ) and a rotational velocity (ω). Coupling between the nearest neighbour spins results in a gradient in the site specific velocity profile.**

Evolution of orientation of the spins follow continuum diffusion equation [Eq.(4)] that is written by combining the continuity equation [Eq.(2)] and the transport equation [Eq.(3)] in the presence of interactions (*U*) between the spins.

$$\frac{\partial c}{\partial t} + \nabla_\theta \cdot j = 0 \qquad (2)$$

where 
$$j = -D_R\left(\nabla_\theta c - \beta c \nabla_\theta U\right) \qquad (3)$$

Therefore, 
$$\frac{\partial c}{\partial t} = \left(D_R \nabla_\theta^2 - \beta \aleph\right) c \qquad (4)$$

Here $c = c(\theta, t)$ is the orientation of spins, *j* is the flux, *U* is the interaction potential given by the Shore – Zwanzig Hamiltonian [Eq.(1)], $D_R$ is the rotational diffusion constant, and $\aleph$ is the torque that is defined as $-\nabla_\theta U$. The above equations determine the orientational evolution for interacting spins in equilibrium conditions. The equations do not account for the effect of the central spin constantly rotating at velocity $\omega_0$. The driving effect of the central spin is not translationally invariant since it depends on the distance of a particular spin from the central spin. Such a macroscopic equation that accounts for the evolution for all the spins, with one of the spins acting as the driving source, is yet to be formulated.

## 3. Simulation Details

We propagate the system of one-dimensional interacting spins by employing rotational Brownian dynamics simuation as shown in Eq.(5).

$$\dot{\theta}_i(t) = (D_R/k_B T)\nabla_\theta U_i(\theta,t) + \sqrt{2D_R}R_i(t) \tag{5}$$

The forces on each spin are given by interactions between the spins written in the Shore-Zwanzig from [Eq.(1)]. The orientation of the spins are evolved following a first order integration scheme [Eq.(5)] resulting in Eq.(6)

$$\theta_i(t+\delta t) = \theta_i(t) + (D_R/k_B T)\nabla_\theta U(\theta,t)dt + \left(\sqrt{2D_R dt}\right)R_i(t) . \tag{6}$$

Here $\theta_i(t)$ is the orientation of the spin at time $t$ and $\theta_i(t+\delta t)$ is the orientation of the spin at time $t+\delta t$. $\nabla_\theta U_i(t)$ is the force on spin $i$ due to nearest neighbour interactions at time $t$.

The coupling constant $J$ given in Eq.(1) is varied to study the effect of increased coupling between the spins. We investigated a system of 101 spins arranged in an one-dimensional array. Each spin interacts with its first nearest neighbor. The central spin rotates at a constant velocity $\omega_0$. Periodic boundary conditions are implemented such that the first spin interacts with the last spin on the array and vice-versa. We use the spin orientation values from 50 independent runs in order to calculate various properties presented in the subsequent sections. The simulations are performed in reduced units. The details of values of the simulation parameters used are as given: dt = 0.1 and $D_R$ = 1.0. $R_i(t)$ is a gaussian distribution with mean 0 and standard deviation 1, scaled by $\sqrt{2D_R dt}$ to represent the thermal noise in the system. The central spin is not subjected to thermal noise in the system. Though the nearest spins interact with the central spin, the dynamics of the central spin itself is unaltered by interactions with nearest spins. Hence, the central spin acts as a source/motor. The evolution of orientation of the central spin is written as

$$\theta_0(t+\delta t) = \theta_0(t) + \omega_0 dt . \tag{7}$$

## 4. Results and discussions
### (A) Phase diagram: Dynamical order and slipping behaviour

Here we explore the two-dimensional parameter space constructed by – (i) the coupling constant, J and (ii) the rotational velocity of the central spin, $\omega_0$. We vary J/k$_B$T from 1.0 to 10.0 and $\omega_0$ from 0.05 to 0.9. For every combination of J and $\omega_0$ we calculate the average rotational velocity (scaled by $\omega_0$) of the entire system and construct a phase diagram shown in **Figure 2**.

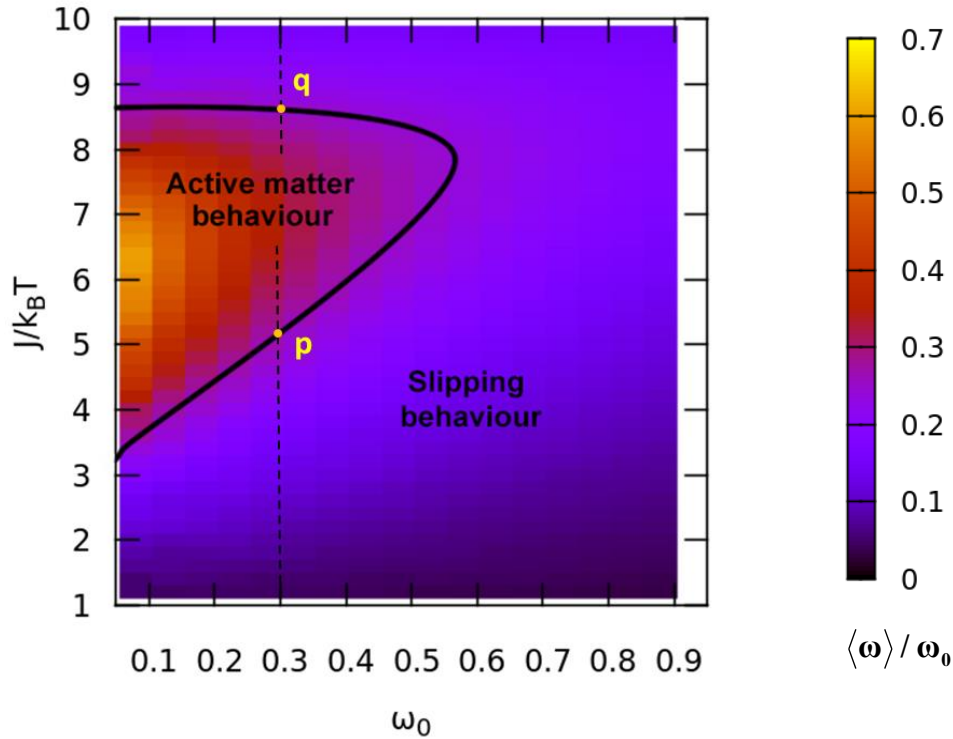

**Figure 2. Phase diagram for a one-dimensional driven Shore-Zwanzig system in J- $\omega_0$ plane. The diagram shows the presence of two distinct regions. In one the system behaves as an active system (yellow/red) and in the other it exhibits a slipping behaviour (violet/purple). The colour bar indicates the scaled (by $\omega_0$) average velocity of the system. When J is varied along a fixed value of $\omega_0$ (shown by the dotted line at $\omega_0$ = 0.3/dt) the system enters into the active matter domain at 'p' and leaves the same at 'q'.**

The phase diagram shows the presence of two distinct regions. For intermediate values of $J/k_BT$ (~4.0 to 8.0) and lower values of $\omega_0$ (less than 0.4), we observe an active matter like behaviour. However, in the limits of large values of J and $\omega_0$ we observe a slipping behaviour. For a large values of *J* and $\omega_0$, the neighbouring spins cannot follow the central sping when it rotates at a high velocity. **Figure 2** *shows that the active behavior is actually observed only in a small region of phase diagram*. This situation can be related to animal/bird flocks. When one object moves extremely fast, others fail to follow it. On the other hand, the unanticipated behaviour in the large *J* limit is hard to comprehend. When the strength of the coupling parameter increases, the non-driven spins influences its neighbours more than that of the driven central spin. Therefore, one might observe the emergence of a chaotic state.

Such phase diagrams can be compared to the order-disorder phase diagrams for equilibrium systems. In a recent study, Klamser *et al.* studied the thermodynamic phases of self-propelled particles in two-dimensions and developed a phase diagram.[43] However, the phase diagram reported in **Figure 2** is different. In our subsequent analyses, we shall concentrate on the active region of the phase diagram.

(B) **Decay of velocity profile**

We perform a detailed study of the emergence of coherence in the velocity of the spins as the interaction between the spins increases. These observations are made by keeping the velocity of the central spin constant at 0.3/step while the interaction strength ($J/k_B T$) is varied between 1.0 to 10.0. The resulting velocity profiles are shown in **Figure 3a** up to J=6.0. This figure embodies one of the main results of the present work. It clearly shows that that in a driven nonequilibrium system, the correlations grow non-linearly with coupling parameter. Note that in an equilibrium system where systems undergo Shore-Zwanzig rotational Brownian motion, correlations in velocity are much shorter ranged. There are of course correlations in orientation, but not in velocity.

The site-specific spin velocities shown are scaled by velocity of the central spin. The correlation lengths are quantified by fitting the velocity profiles to an exponential or bi-exponential decay followed by the extraction of the average correlation length, $\langle \xi \rangle$ [**Table 1**]. Similar to the observations made with Vicsek model, where decreased noise and increased system density gives rise of large correlation in the velocities of the spins, we find an emerging coherence in the velocity of the spins with increased coupling interactions among the spins. However, beyond a certain value of J (here J > 6), we again observe short ranged spatial correlations (that is, a decrease in $\langle \xi \rangle$) and an exponential decay of the velocity profile [**Figure 3b**]. The non-monotinicity of **Figure 3b** originates because of two competing factors, namely, (i) the influence of the central driven spin which leads to a dynamically ordered state and (ii) the influcne of the non-rotating spins on its neighbours which leads to a chaotic disordered state. Sharpness near the peak in the variation of correlation length with J at that givent $\omega_0$ is due to the crossing the two boundaries (in and out) of the phase diagram.

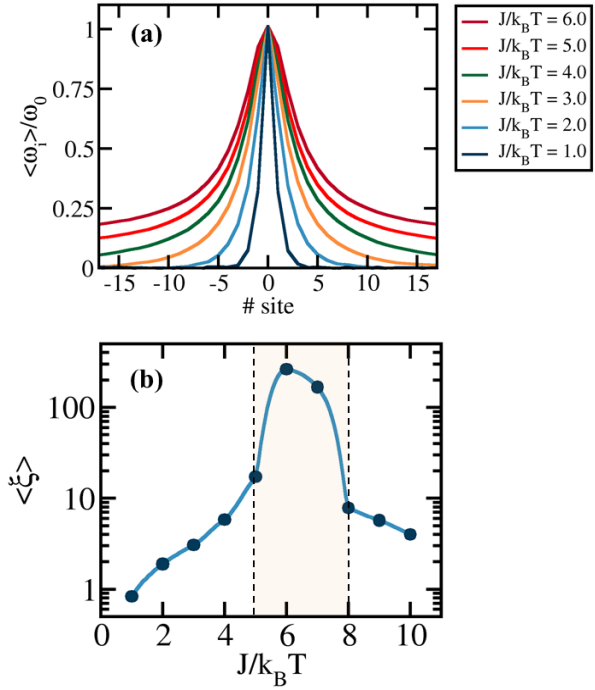

**Figure 3. (a)** Average velocity of the spins scaled by velocity of the central spin as a function of position of the spin from the central spin. The average velocities are obtained by averaging over 50 independent simulations. Different colours show the velocity profiles for different coupling strength between the spins (J) as indicated in the figure legend. As the coupling between the spin increases, the correlation length of velocity of the spins increases. The simulations were performed on a system size of 101 spins and the central spin is positioned at site 0. As the coupling strength is varied, the velocity of the central spin is kept constant at $\omega_0$=0.3/dt. The data are fitted to an exponentially/biexponentially decaying function as given in Table 1. **(b)** Variation of the average correlation length ($\langle\xi\rangle$) with the coupling constant J shows an initial rapid increase followed by a decrease for J > 6. The shaded region shows the active (large spatial correlation) region which is also shown qualitatively by 'p' and 'q' in Figure 2.

**Table 1.** Exponential/bi-exponential fitting parameters for the rotational velocity decay profiles shown in Figure 3.

| $J/k_BT$ | $\xi_1$ | $\xi_2$ | $\langle\xi\rangle$ |
|---|---|---|---|
| 1.0 | 0.83 (100%) | --- | 0.83 |
| 2.0 | 1.89 (100%) | --- | 1.89 |
| 3.0 | 3.05 (100%) | --- | 3.05 |
| 4.0 | 3.65 (92%) | 30.6 (8%) | 5.8 |
| 5.0 | 4.02 (88%) | 211.4 (12%) | 28.8 |
| 6.0 | 4.5 (84%) | 1613.6 (16%) | 261.9 |
| 7.0 | 6.0 (86%) | 1158.0 (14%) | 167.0 |
| 8.0 | 7.8 (100%) | --- | 7.8 |
| 9.0 | 5.7 (100%) | --- | 5.7 |
| 10.0 | 4.0 (100%) | --- | 4.0 |

### (C) Effect of interaction strength on the velocity distributions

We turn to the probability distribution of velocities as a means of qualitatively understanding the entropy of the spin system under different coupling strengths. A broader distribution of the velocities would indicate higher variability in the velocities of the individual spins and hence more entropy of the system.

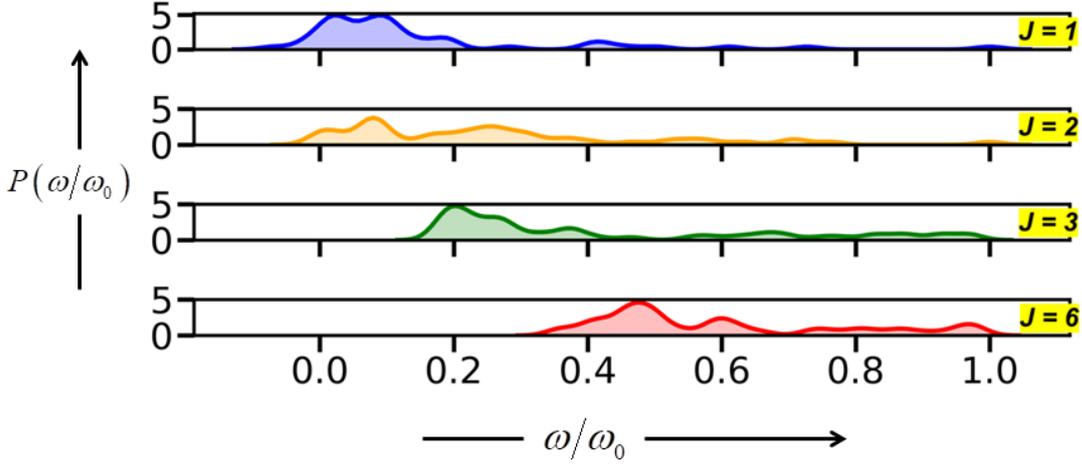

**Figure 4. Normalised probability distribution of scaled velocity of the spins. The velocities are scaled by velocity of the central spin. As the coupling between the spins increases, the distribution becomes narrower and the mean velocity increases. The results correspond to simulations performed on a 101 spin system with the central spin (#0) rotating at $\omega_0 = 0.3/dt$.**

**Figure 4** shows the normalised probability distribution of the spin velocitites as a function of the coupling strength of between the spins. We observe that as the coupling between the spin increases, the mean velocity shifts to higher values, closer to the velocity of the central spin and the distribution becomes narrower. This is interpreted as the decreasing entropy of the system with increased coupling between the spins.

**(D) Effect of the velocity of the central spin**

The central spin, rotating at a constant velocity, produces the motor force for the system. Hence the rotational velocity of the central spin dictates the dynamics of all the other spins coupled to the central spin. We observe that at a given coupling strength, as the velocity of the central spin increases, the average velocity of the spins decreases. Further systematic investigation on the relationship between velocity of the driver spin and coupling between the spins is yet to be done to ascertain the nature of transition from the decorelated state to a state of correlation in the velocity of the spin.

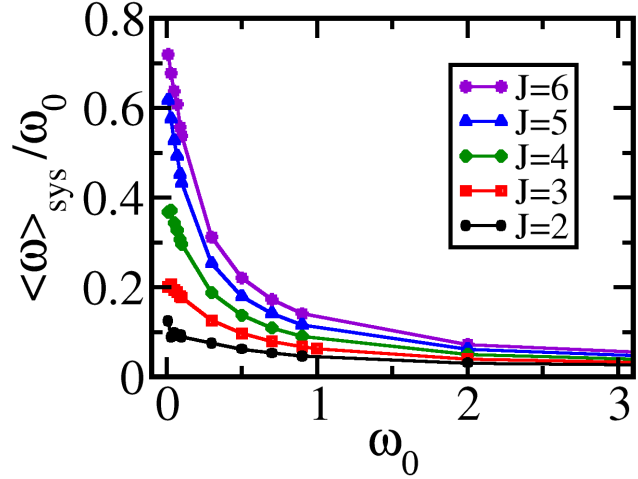

**Figure 5.** Average velocity of the spins scaled by velocity of the central spin as a function of the velocity of the central spin ($\omega_0$). The average velocities are obtained by averaging over 50 independent simulations and averaged across the left and right of the central spin for each simulation. As the velocity of the central spin increases, the correlation length of velocity of the spins decreases. The simulations were performed on a system size of 101 spins. The central spin is positioned at site 0. As the velocity of the central spin is varied from 0 to 3, the coupling constant (J) is also varied from 2 to 6.

We fit the data shown in **Figure 5** with a stretched exponential function: $y = a_0 \exp\left[-\left(x^\beta / \alpha\right)\right]$, and provide the fitting parameters in **Table 2**.

**Table 2.** Fitting parameters for the data shown in Figure 7 with a stretched exponential function: $y = a_0 \exp\left[-\left(x^\beta / \alpha\right)\right]$.

| $J / k_B T$ | $a_0$ | $\beta$ | $\alpha$ |
|---|---|---|---|
| 2.0 | 0.17 | 0.27 | 0.78 |
| 3.0 | 0.25 | 0.49 | 0.79 |
| 4.0 | 0.43 | 0.61 | 0.62 |
| 5.0 | 0.74 | 0.57 | 0.52 |
| 6.0 | 0.82 | 0.65 | 0.50 |

### (E) Correlations in a steady state

Since the time of Onsager and Prigogine, the nature of stability and dynamics (decay of fluctuations) of systems out of equilibrium has drawn tremendous interest and is a subject of fundamental interest. Therefore, we have attempted to characterize the decay of fluctuations in such a driven system in a steady state far from equilibrium. We have studied the decay of the orientational time correlation function. The total orientational time correlation function in equilibrium systems is a measure of total dipole moment time correlation of the system. This was the quantity studied by Shore and Zwanzig, and is routinely studied in dielectric relaxation experiments, giving rise to frequency dependent dielectric function, ε(ω). The total dipole moment is given by

$$M = \sum_i \cos\theta_i \tag{8}$$

At equilibrium the fluctuations of total orientation $M$ of the spins follow linear response theory. The normalised time correlation of $M$ given as $<M(t)M(0)>/<M(0)^2>$ decays nearly exponentially, with a time constant given by $1/2D$. It is therefore of fundamental interest to probe the decay of the TCF in the system far from equilibrium.

The system studied here has an inherent heterogeneous nature. The symmetry of the system and the driving central spin introduces a position dependence in the dynamics. We investigate the orientation time correlation of the spins for $J/k_BT=3.0$. The orientation time correlation of the fifth spin from the central spin is shown in **Figure 6**. We observe that at a coupling strength of $J/k_BT=3.0$, the decay of orientation time correlation of the third and fifth spins are faster than that of the non-driven system. At lower coupling strength, it is observed that the spins lag behind the central spin.

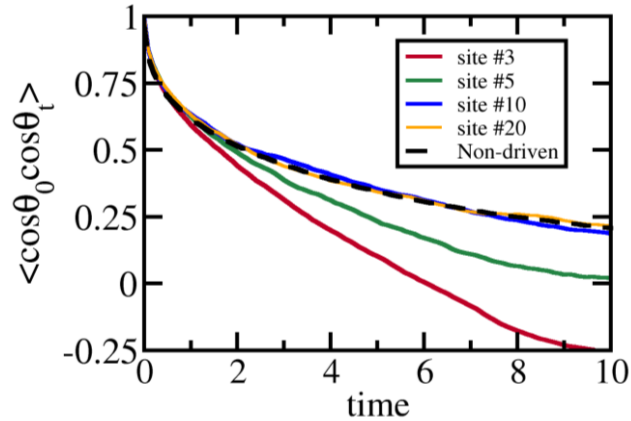

**Figure 6.** Site specific orientation time correlation (OTC) function for the driven system and non-driven system (dashed line). The OTC for the third, fifth, tenth, and twentieth spins from the central spin are shown for the driven systems. The OTCs are shown for $J = 3.0$ and $\omega_0 = 0.3/dt$. Due to the homogeniety in the non-driven system, orientation correlation is averaged across the different spins (shown by dashed line).

The decay of normalized time correlation function of $M$ is given as:

$$\frac{\langle M(0)M(t)\rangle}{\langle M(0)^2\rangle} = \exp[-Bt] \tag{9}$$

Here $B = \mu\omega^2/\zeta$, where $\zeta$ is the friction coeffient. Using the Einstein's relation between friction coeffcient ($\zeta$) and diffusion constant ($D$), $B$ can be re-written as $B = D\mu\omega^2/k_B T$. Thus as the coupling between the spins increases (or, the velocity of the central spin increases) the time correlation of *M* decays faster. We note here that the assumption of expotential decay of time correlation function of M need not be valid for the driven non-equilibrium system at steady state.

## 5. Conclusion

Here we study a simple model of an interacting system to capture some aspects of a driven system which is the hallmark of an active matter. To the best of our knowledge, this could be one of the first studies that consider rotational motion to mimic the emergence of order in driven systems. We employed the elegant Shore-Zwanzig Hamiltonian and Brownian dynamics simulations to examine the emergence of dynamical correlations among the spins in the system. We find several potentially interesting results, like growth in dynamical correlations as the coupling between the spins increases, and/or the amplitude of the velocity is changed.

The main result is the emergence of a coherent state at certain range of values of the coupling parameter J and the driving rotational velocity $\omega_0$ of the central spin. It can be easily understood that a single active spin at center cannot create a macroscopic dynamically ordered state. This is because the dynamical correlation at fixed J and temperature has a finite length. However, what makes sense is to consider a finite concentration of active spin rotors. In that situation, a dynamically ordered state shall exist at a spin concentration beyond certain value which shall be determined by the spin correlation length. The interesting question that arises in such a situation is that the active rotors spins can interact with each other dynamically through other spins, and this can create a rotor phase, like the one reported here. If the dynamical correlation length is $\xi$ for an isolated active spin, then simple mean-field argument shall give the concentration required to be more than $N/\xi$, where N is the total number of spins. This concentration should actually be less than $N/\xi$, and precise determination is yet to be made. Perhaps more interesting case arises when the active molecules are placed randomly along the lattice. That is, the spins in the lattice are selected to be active in a random fashion. The other interesting case obviously is the two dimensional active spins. Here the XY model shows a phase transition. The interaction between phase transition and activity can be of considerable interest, especially near the critical point.

We find a slipping behavior when the angular velocity is large and no amount of coupling can give rise to the coherence in motion among the spins. The present model differs from the existing models of active matter in an important aspect. Here only the central spin is self-propelled. Thus, while the onset of a collective motion due to interaction was expected, we find several unexpected results also. What was not so obvious is the slipping motion in the limits of large interaction and/or large rotational velocity. Also, the emergence of the bi-exponential decay of the velocity time correlation function and the non-monotonic behavior of the correlation length were unexpected. While it is easy to understand the crossover to non-active or slipping region of the system with increase of $\omega_0$, the occurance of the same with increase of J at constant $\omega_0$ is hard to understand.

Although the model studied here is different from the model of Vicsek *et al.* and the model of del Junco *et al.*, there are also certain similarities. The emergence of the coherent state due to nearest neighbour interaction is reminiscent of Vicsek model,[1] but here the interaction is real and originates from a Hamiltonian. The emergence of a phase transition like scenario and the change in velocity distribution the coherent state is similar to that of the del Junco *et al*.[28] We believe that this model can be greatly generalized to study many phenomena. We hope to pursue this approach in future work.

**Acknowledgement**

We thank Prof. Chandan Dasgupta (Department of Physics, IISc) for several useful discussons and suggestions. BB thanks the National science chair professorship (DST-SERB), India for financial support. SM thanks the same for a Research Associateship. SSI is thankful to the Ministry of Human Resource Development (MHRD), India for a research fellowship.